%% file: main.tex
\documentclass{article} 
\usepackage{iclr2017_conference,times}
\usepackage{hyperref}
\usepackage{mathtools}
\usepackage{graphicx}
\usepackage{subfig}
\usepackage{bbm}
\usepackage{url}
\usepackage{tikz}
\usetikzlibrary{arrows,shapes,trees,backgrounds,positioning}

\title{Song From PI: A Musically Plausible Network For Pop Music Generation}

\author{Hang Chu, Raquel Urtasun, Sanja Fidler\\
Department of Computer Science\\
University of Toronto\\
Ontario, ON M5S 3G4, Canada \\
\texttt{\{chuhang1122,urtasun,fidler\}@cs.toronto.edu} \\
}

\begin{document}

\maketitle

\input{iclr-abstract}

\input{iclr-introduction}
\input{iclr-related}

\input{iclr-music}
\input{iclr-method}
\input{iclr-experiment}
\input{iclr-application}
\bibliography{iclr2017_conference}
\bibliographystyle{iclr2017_conference}

\end{document}

%% file: iclr-abstract.tex
\begin{abstract}
We present a novel framework for generating pop music. Our model is a hierarchical Recurrent Neural Network, where the layers and the structure of the hierarchy encode our prior knowledge about how pop music is composed. In particular, 
the bottom layers generate the melody, while the higher levels produce the drums and chords. We conduct several human studies that show strong preference of our generated music over that produced by the recent method by Google. We additionally show two applications of our framework: neural dancing and karaoke, as well as neural story singing. 
\end{abstract}


%% file: iclr-introduction.tex
\section{Introduction}

Neural networks have revolutionized many fields. They have not only proven to be powerful in performing perception tasks such as image classification and language understanding, but have also shown to be surprisingly good ``artists''. In~\cite{Gatys16}, photos were turned into paintings  by exploiting particular drawing styles such as Van Gogh's,~\cite{kiros2015skip} produced stories about images biased by writing style (e.g., romance books),~\cite{Karpathy16} wrote Shakespeare inspired novels, and~\cite{SimoCVPR15} gave fashion advice.

Music composition is another artistic domain where neural based approaches have been proposed.  Early approaches exploiting  Recurrent Neural Networks (\cite{Bharucha89,Mozer96,Chen01,eck2002first}) date back to the 80's. The main variations between the different models is the representation of the notes and  the  outputs they produced, which  typically encode melody and chord. Most of these approaches were single track, in that they produced only one note per time step.
The exception is~\cite{boulanger2012modeling} which generated polyphonic music, i.e., simultaneous independent melodies. 

In this paper, we aim to generate pop music, where the melody but also chords and other instruments make up what is typically called a song. 
We draw inspiration from the Song from $\pi$ by~\cite{songfrompi}~\footnote{\url{https://youtu.be/OMq9he-5HUU}}, a piano video on Youtube, where the pleasing music is created from a sequence of digits of $\pi$. This video shows both the randomness and the regularity of music. On one hand, since any possible digit sequence is a subset of the $\pi$ digit sequence, this implies that pleasing music can be created even from a totally random base signal. On the other hand, the composer uses specific rules such as \textit{A Harmonic Minor} scale and \textit{harmonies} to convert the digit sequence into a music sheet. It is these rules that play the key role in converting randomness into music.

Following the ideas of Songs from $\pi$, we aim to generate both the melody as well as accompanying effects such as chords and drums. Arguably, these turn even a not particularly pleasing melody into a well sounding song. We propose a hierarchical approach, where each level is a Recurrent Neural Network producing a key aspect of the song. The bottom layers generate the melody, while the higher levels produce drums and chords. This enables the drum and chord layers to compensate for the melody in order to produce appleasing music. Adopting the key idea from Songs from $\pi$, we condition our model on the scale type allowing the melody generator to learn the notes that are typically played in a particular scale.


We train our model on 100 hours of midi music containing user-composed pop songs and video game music. We conduct human studies with music generated with our approach and compare it against a recent approach by Google, showing that our songs are strongly preferred over the baseline. In our human study we also perform an ablation analysis of our model.  We additionally show two new applications: neural dancing and karaoke as well as neural music singing. As part of the first application we generate a stickman dancing to our music and lyrics that can be sung with, while in the second application we condition on the output of~\cite{kiros2015skip} which writes a story about an image and convert it into a pop song.
We refer the reader to {\color{magenta} http://www.cs.toronto.edu/songfrompi/ }for our demos and results.

%% file: iclr-related.tex
\section{Related Work}

Generating music has been an active research area for decades. It brings together machines learning researchers that aim to capture the complex structure of music (\cite{eck2002first,boulanger2012modeling}), as well as music professionals (\cite{Chan06}) and enthusiasts (\cite{Johnson,DeepHear}) that want to see how far a computer can get to be a real composer. Real-time music generation is also explored for gaming (\cite{Engels15}).

Early approaches mostly instilled knowledge from  music theory into generation, by using rules of how music segments can be stitched together in a plausible way, e.g., \cite{Chan06}. On the other hand, neural networks have been used for music generation since the 80's (\cite{Bharucha89,Mozer96,Chen01,eck2002first}). ~\cite{Mozer96} used a Recurrent Neural Network that produced pitch, duration and chord at each time step. Unlike most other neural network approaches, this work encodes music knowledge into the representation. ~\cite{eck2002first} was first to use LSTMs to generate both  melody and chord. Compared to~\cite{Mozer96}, the LSTM captured more global music structure across the song.

Like us,~\cite{Kang2012} built upon the randomness of melody by trying to accompany it with drums. However, in their model the scale type is enforced. No details about the model are presented, and thus it is virtually impossible to compare to. \cite{boulanger2012modeling} propose to learn complex polyphonic musical structure which has multiple notes playing in parallel through the song. The model is single-track in that it only produces melody, whereas in our work we aim to produce multi-track songs. Just recently,~\cite{huang2016deep} proposed a 2-layer LSTM that, like~\cite{boulanger2012modeling},  produces music that is more complex than a single note sequence, and is able to produce chords.
The main novelty of our work over existing approaches is a hierarchical model that  incorporates knowledge from music theory to build the neural architecture, and produces multi-track pop music (melody, chord and drum). We also present two novel fun applications. 






%% file: iclr-music.tex

\section{Concepts from Music Theory}

We start by introducing the basic notation and definitions from music theory.
A {\bf note} defines the basic unit that music is composed of. Music follows the {\bf 12-tone} system, i.e., 12 is the cycle length of all notes. 
The 12 tones are: $C$, $C^\sharp/D^\flat$, $D$, $D^\sharp/E^\flat$, $E$, $F$, $F^\sharp/G^\flat$, $G$, $G^\sharp/A^\flat$, $A$, $A^\sharp/B^\flat$, $B$. A {\bf bar} is a short segment of time that corresponds to a specific number of beats (notes). The boundaries of the bar are indicated by vertical bar lines. 

\textbf{Scale} is a subset of notes. There are four types of scales most commonly used: \textit{Major} (\textit{Minor}), \textit{Harmonic Minor}, \textit{Melodic Minor} and \textit{Blues}. Each scale type specifies a sequence of relative \emph{intervals} (or shifts) which act relative to the   starting note. For example, the sequence for  the scale type \textit{Major} is $2 \rightarrow 2 \rightarrow 1 \rightarrow 2 \rightarrow 2 \rightarrow 2 \rightarrow 1$. Thus,  \textit{C Major} specifies the starting note to be C, and applying the relative sequence of shifts yields:  $C \xrightarrow{2} D \xrightarrow{2} E \xrightarrow{1} F \xrightarrow{2} G \xrightarrow{2} A \xrightarrow{2} B \xrightarrow{1} C$. 
The subset of notes specified by \textit{C Major}  is thus C, D, E, F, G, A, and B (a subset of seven notes). 
All scales types have a subset of seven notes except for Blues which has six. 
In total we have 48 unique scales, i.e. 4 scale types and 12 possible starting notes. We treat \textit{Major} and \textit{Minor} as one type as 
for a \textit{Major} scale there is always a \textit{Minor} that has exactly the same set of notes. In music theory, this is referred to as \textit{Relative Minor}.

\textbf{Chord} is a group of notes that sound good together. Similarly to scale, a chord has a start note and a type defining a set of intervals. There are mainly 6 types in triads chords: \textit{Major Chord}, \textit{Minor Chord}, \textit{Augmented Chord}, \textit{Diminished Chord}, \textit{Suspended 2nd Chord}, and \textit{Suspended 4th Chord}.

\textbf{The Circle of Fifths} is often used to produce a chord progression. It maps 12 chord starting notes to a circle. When changing from one chord to another chord, moving to a nearby chord on the circle is often preferred as this forms a \emph{strong chord progression} that produces the sense of harmony. 


%% file: iclr-method.tex
\section{Hierarchical Recurrent Networks for Pop Music Generation}

We follow the high level idea behind the Song from $\pi$ to define our model. In particular, we generate music with a hierarchical Recurrent Neural Network where the layers and the structure of the hierarchy encode our prior knowledge about how pop music is composed. We first outline the model and describe the details and justifications for our choices in the subsections that follow. 

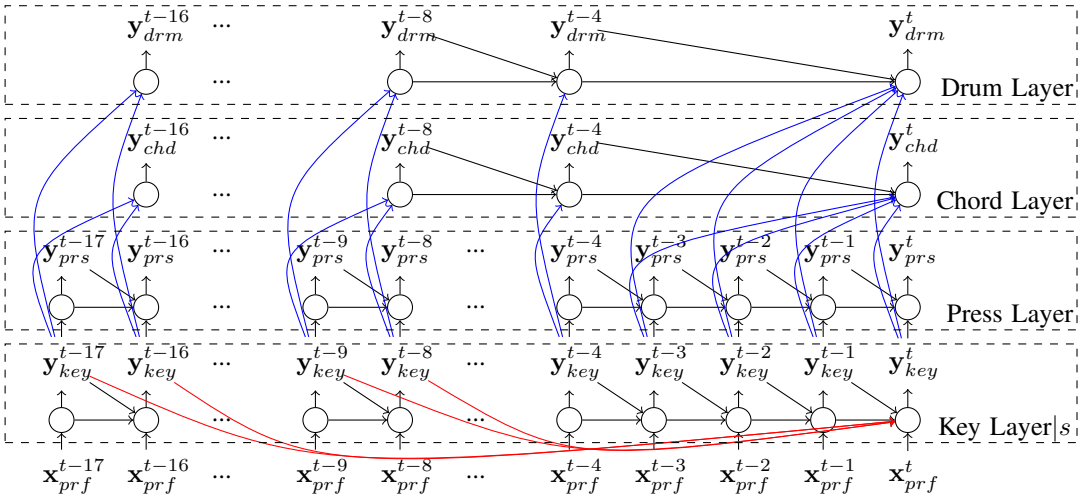
\begin{figure}[t!]%
\centering
\input{fig1}
\caption{\small Overview of our framework. Only skip connections for the current time step $t$ are plotted.}%
\label{fig:overview}
\end{figure}

We condition our generation on the scale type, as this helps the model to pick up the regularities in pop songs. We encode melody with two random variables at each time step, representing which key is being played (the \emph{key layer}) and the duration that the key will be pressed (the \emph{press layer}). The melody is generated conditioned on the scale, which does not vary across the song as is typically the case in pop music. 
We assume the drums and the chords are independent given the melody.
Thus conditioned on the melody, at each time step we generate the chord (the \emph{chord layer}) as well as the drums (the \emph{drum layer}). The output at all layers yields the final song. We refer the reader to Fig.~\ref{fig:overview} for an illustration of our hierarchical model.

\subsection{The role of  Scale}

It is known from music theory that while in principle  each song has 12 tones to choose from, most of the notes are in fact only using the  six (for Blues) or seven (for other scales) tone subsets specified by the scale rule. We found that by conditioning the music generator on scale  it captures these regularities more easily. However, we do not enforce the notes to be generated from the  subset and allow our model to generate notes outside the scale.

\begin{figure}[t!]%
\centering
\addtolength{\tabcolsep}{-8pt}
\begin{tabular}{cccc}
\includegraphics[width=0.235\linewidth]{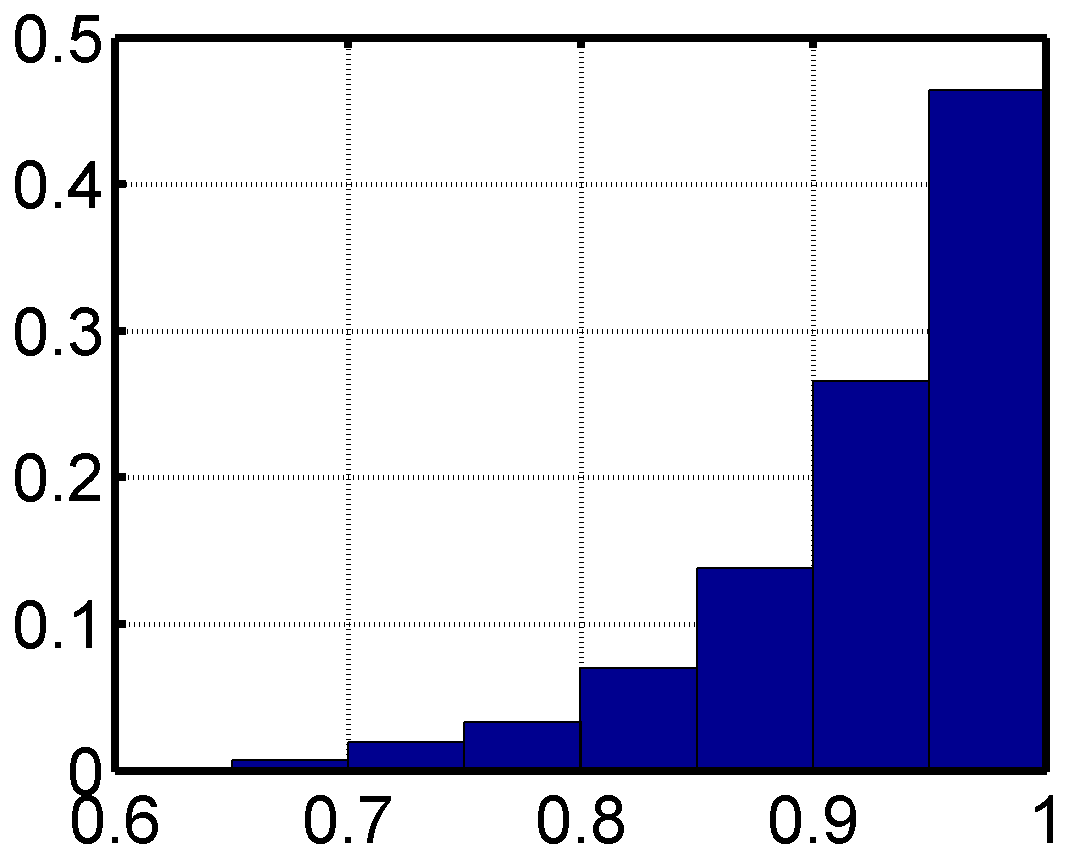}$\ \,$ & 
\includegraphics[width=0.235\linewidth]{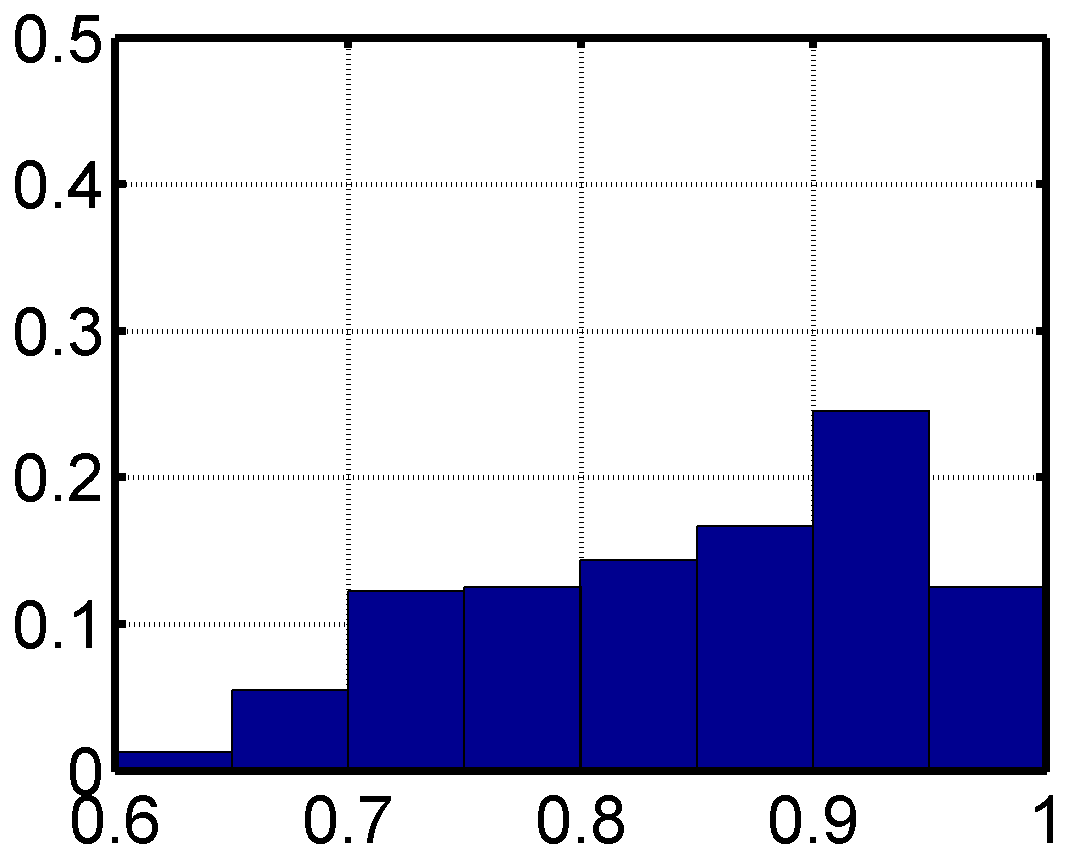}$\ \,$ &
\includegraphics[width=0.235\linewidth]{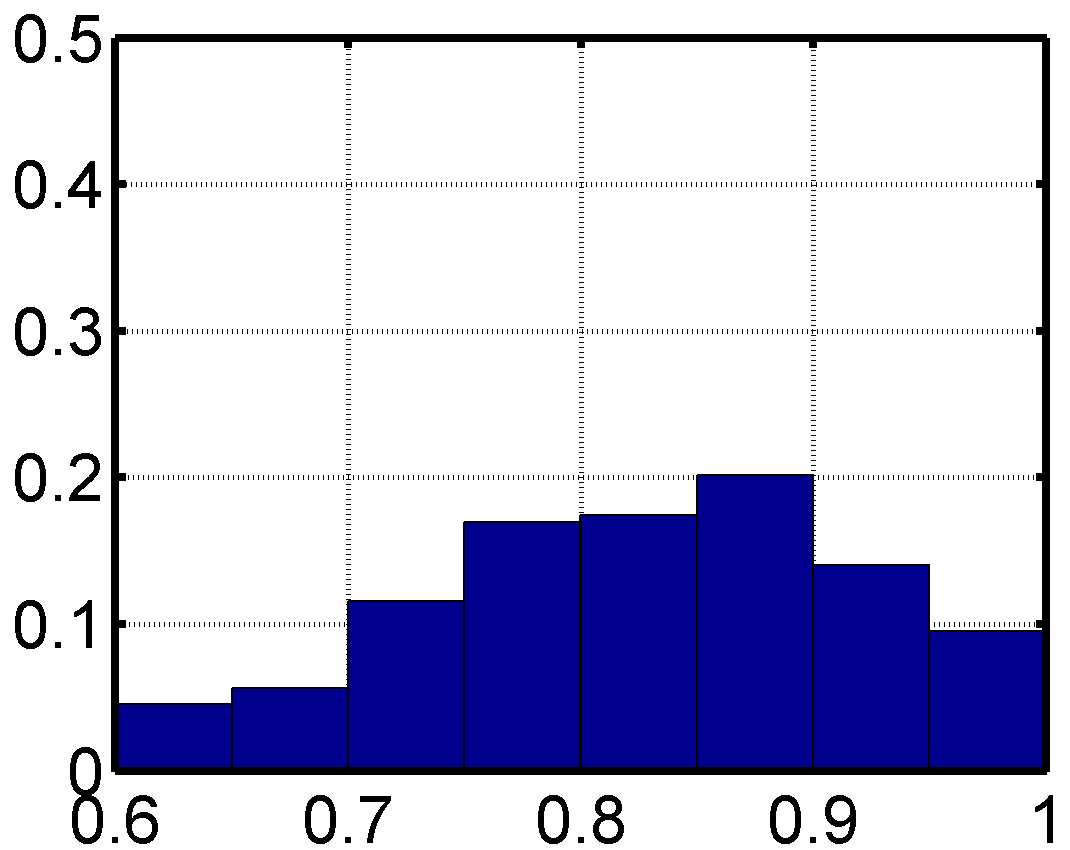}$\ \,$ &
\includegraphics[width=0.235\linewidth]{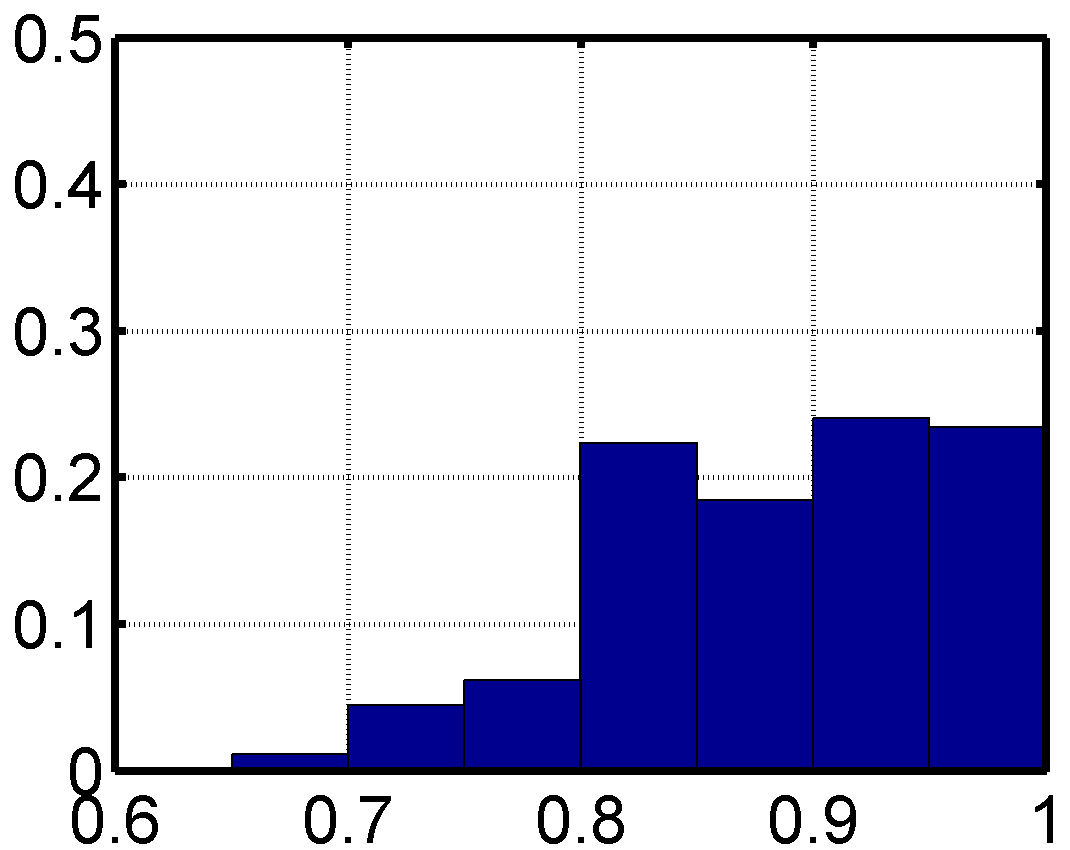}\\
(a) & (b) & (c) & (d) \\
\end{tabular}
\caption{\small Distribution of within-scale note ratio for four scale types. x-axis: percentage of tones within the scale type's tone set, y-axis: percentage of songs of the scale type. (a)-(d) shows \textit{Major}(\textit{Minor}), \textit{Harmonic Minor}, \textit{Melodic Minor}, and \textit{Blues}, respectively.}%
\label{fig:scale} %
\end{figure}

We confirm the above musical fact by analysing over 100 hours of pop song music from the~\cite{midiman} dataset. Since scale is defined relative to a starting note, we first try to factor out its influence and normalize all songs to have identical start note.
To identify the scale of a song, we compute the histogram over the 12 tones and match it with the 48 tone subsets of 4 scale types with 12 different start notes.
We then normalize all songs to have start note $C$ by applying a constant shift on all notes. This allows us to categorize any song into 4 scale types. 
Since this shift affects all notes at once, it does not affect how the song sounds (its harmony). 
Our analysis shows that for all notes in all \textit{Major} scale songs, $94.66\%$ are within the tone subset. For \textit{Harmonic Minor}, \textit{Melodic Minor}, and \textit{Blues} the percentage of notes that belong to the main tone set is $87.16\%$, $85.11\%$, and $90.93\%$, respectively. We refer the reader to Fig.~\ref{fig:scale}, where the x-axis denotes the percentage of within-scale notes of a song, and the y-axis indicates how many songs in the dataset have that percentage. Note that the majority of the notes follow the scale rule. Furthermore, different scale types have different inlier distribution. 
We thus represent scale with a single random variable $s\in \{1, \cdots, 4\}$ which is fixed for the whole song, and condition the model on it. \footnote{For readers with musical background, the \textit{Twelve-Tone Serialism} technique~\cite{schoenberg1951style} prevents  emphasis of any one tone. However, our data analysis indicates that pop music is not influenced by it.}

\subsection{Two-layer RNN for Melody Generation}

We represent the melody with two random variables per time step: which key is pressed, and the duration of the press. A Recurrent Neural Network (RNN) is used to generate the key condition on the scale. Then conditioned on the output of the key layer, a second RNN generates the duration of the press at each time step.

In this paper we take advantage of LSTMs, which in their most basic form (single layer) compute the hidden state $h^t$ given the input $\mathbf{x^t}$ by
\begin{align}\label{lstm}
  f^t &= \sigma(W_f[\mathbf{x^t},h^{t-1}]) \\ 
  i^t &= \sigma(W_i[\mathbf{x^t},h^{t-1}]) \nonumber\\
  o^t &= \sigma(W_o[\mathbf{x^t},h^{t-1}]) \nonumber\\
  \widetilde{C^t} &= tanh(W_C[\mathbf{x^t},h^{t-1}]) \nonumber\\
  C^t &= f^t \odot C^{t-1}+i^t \odot \widetilde{C^t} \nonumber\\
  h^t &= o^t \odot tanh(C^t) \nonumber
\end{align}
with $W_f, W_i, W_o, W_C$ learnable parameters. Here 
 $f$, $i$, $o$, $C$, $\widetilde{C}$, and $h$ denote the forget gate, input gate, output gate, cell state, input cell state and hidden state. 

In particular, we model the key layer with a two-layer LSTM with 512-dimensional hidden state, which outputs a note (key) at each time step. 
Note that we condition on scale $s$, thus we have different parameters per scale. 
We only allow notes between $C3$ to $C6$ as notes outside this range are usually too low or too high to sound good. 
We remind the reader that given a scale, seven (or six for blues) out of the twelve notes (per octave) are statistically more plausible, however we allow the model to choose from all 12. This results in a 37-dimensional output, as there are 36 possible notes corresponding to 3 octaves with 12 notes per octave, plus silence. 
Let $h^t_{key}$ be the hidden state of the second key decoder layer at time $t$.  We compute the probability of each key using the softmax:
\begin{equation}
P({\bf y}^t_{key}) \propto \text{exp} ( {\bf v}_{{\bf y}^t_{key}} h^t_{key} )
\end{equation}
where ${\bf v}_{{\bf y}^t_{key}}$ is the row of ${\bf V}$ (the output embedding matrix of notes), corresponding to note ${\bf y}^t_{key}$. 

As input to the LSTM we  use a vector that concatenates multiple features: a one-hot encoding of the previous generated note ${\bf y}^{t-1}_{key}$, Lookback features, and the melody profile.   
The Lookback features were proposed by Google Magenta (\cite{magenta}) to make it easier for the model to memorize recently produced notes and potentially repeat them. They include skip connections from two and one bar ago (a bar is 8 consecutively played notes), i.e., ${\bf y}^{t-16}_{key}$ and ${\bf y}^{t-8}_{key}$. 
They also contain two additional features, indicating whether the last generated key has been copied from one or two bars ago, i.e. $\mathbbm{1}({\bf y}^{t-1}_{key},{\bf y}^{t-1-8}_{key})$ and $\mathbbm{1}({\bf y}^{t-1}_{key},{\bf y}^{t-1-16}_{key})$.
They also add a 5-dimensional feature indicating a binary encoding of the current time $t$.
This helps the model keep track where in a $4-$bar range it is, and thus produce music accordingly.

In addition, we introduce a new feature which we refer to as the \emph{melody profile}. Intuitively, the profile represents the high-level music flow. To get the profile for each  song,  we compute the local note histogram at each time step with width of two bars, and cluster all local histograms within the song into 10 clusters via k-means. We order the 10 clusters with mean note ordered from low to high as cluster 1 to 10, and apply moving averages on the cluster id sequence to encourage local smoothness. This results in a 10-dimensional one-hot vector representation of the cluster id for each time step.
This additional information allows the user to set the melody's ups and downs of the song.

The keys alone are not sufficient to describe how the melody is performed.  
Additionally we also need to know the duration that each key needs to be pressed for. Towards this goal, conditioned on the melody, we generate the duration of each key with a two-layer LSTM with a 512-dimensional hidden state. 
We represent the duration of pressing as a forward counting sequence that is conditioned on the generated melody. The press outputs 1 when a new key is pressed, and sequentially outputs 2, 3, 4 and so on as the key is held on. When the current key is released, the press counter is reset to 1. 
Compared to the event on-off representation of \cite{magenta}, our representation learns the melody flow and how to press separately. This is important, as  \cite{magenta} has extremely unbalanced output distributions dominated by the repeat-of-holding event. We represent press ${\bf y}^t_{prs}$ as a 8-dimensional one-hot vector. 
The input to our LSTM is ${\bf y}^{t-1}_{prs}$, concatenated with the 37-dimensional one-hot encoding of the melody key ${\bf y}^{t}_{key}$.




\subsection{Chord and Drum RNN Layers}

\begin{figure}[t!]%
\centering
\addtolength{\tabcolsep}{-8pt}
\begin{tabular}{cccc}
\includegraphics[width=0.235\linewidth]{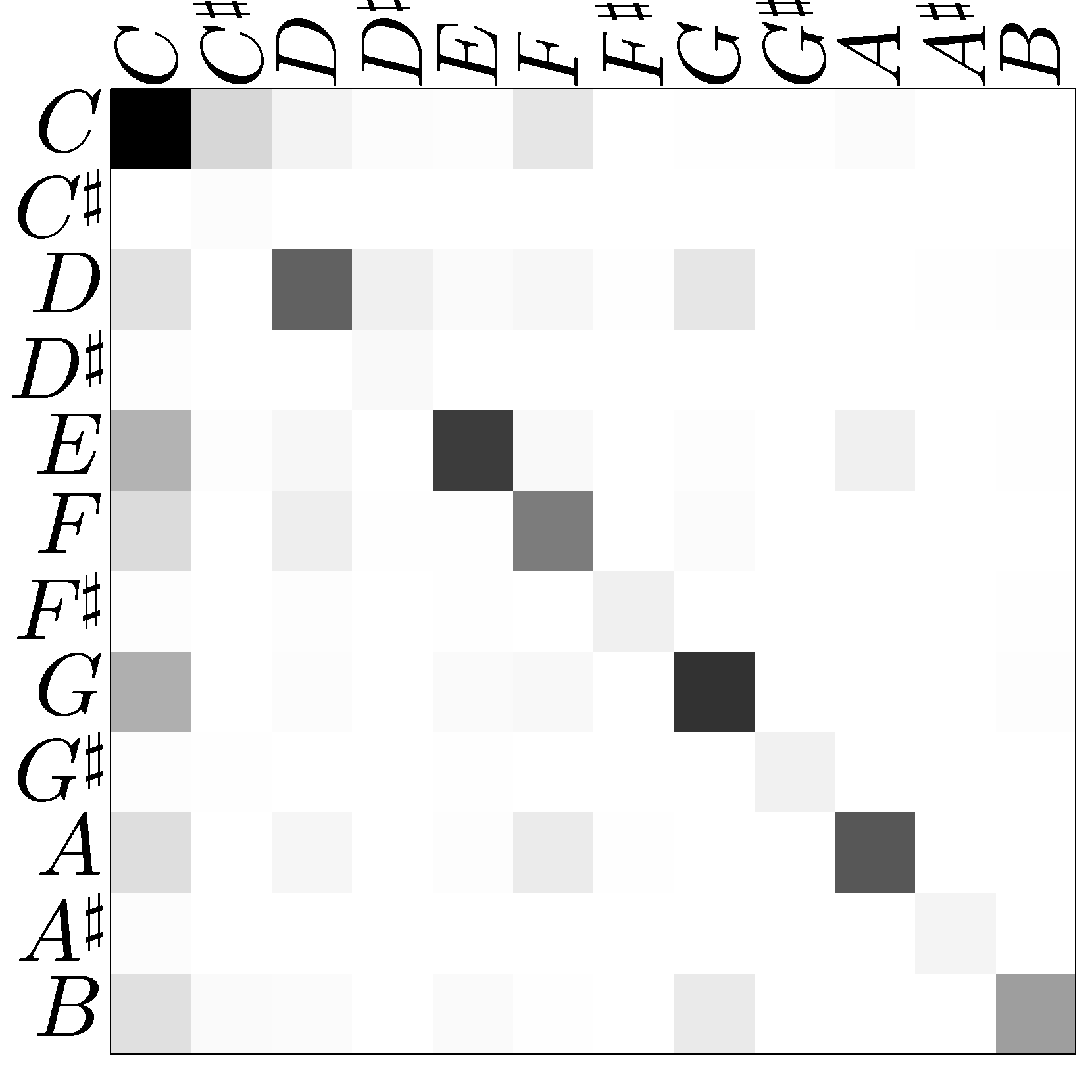}$\ \,$ & 
\includegraphics[width=0.235\linewidth]{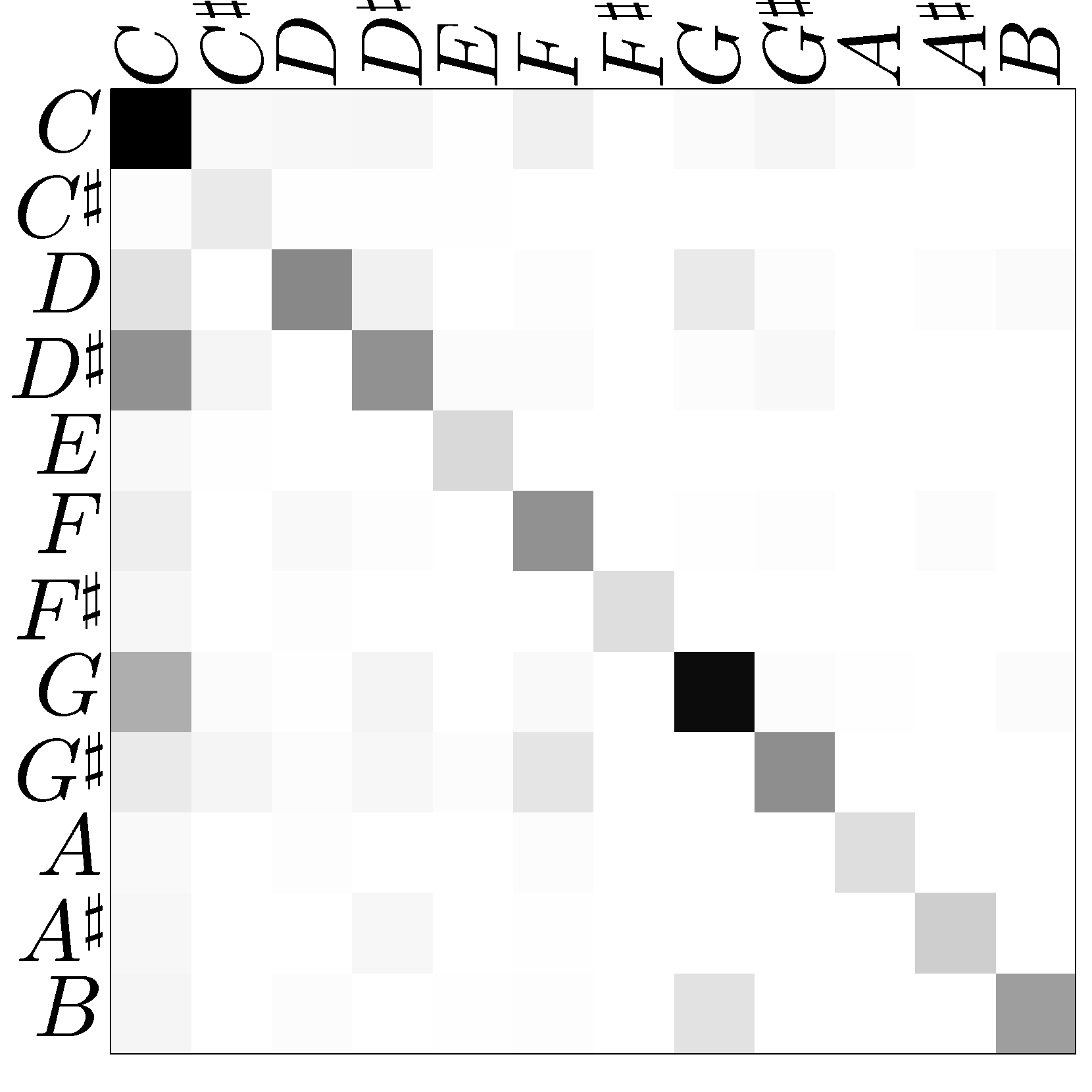}$\ \,$ &
\includegraphics[width=0.235\linewidth]{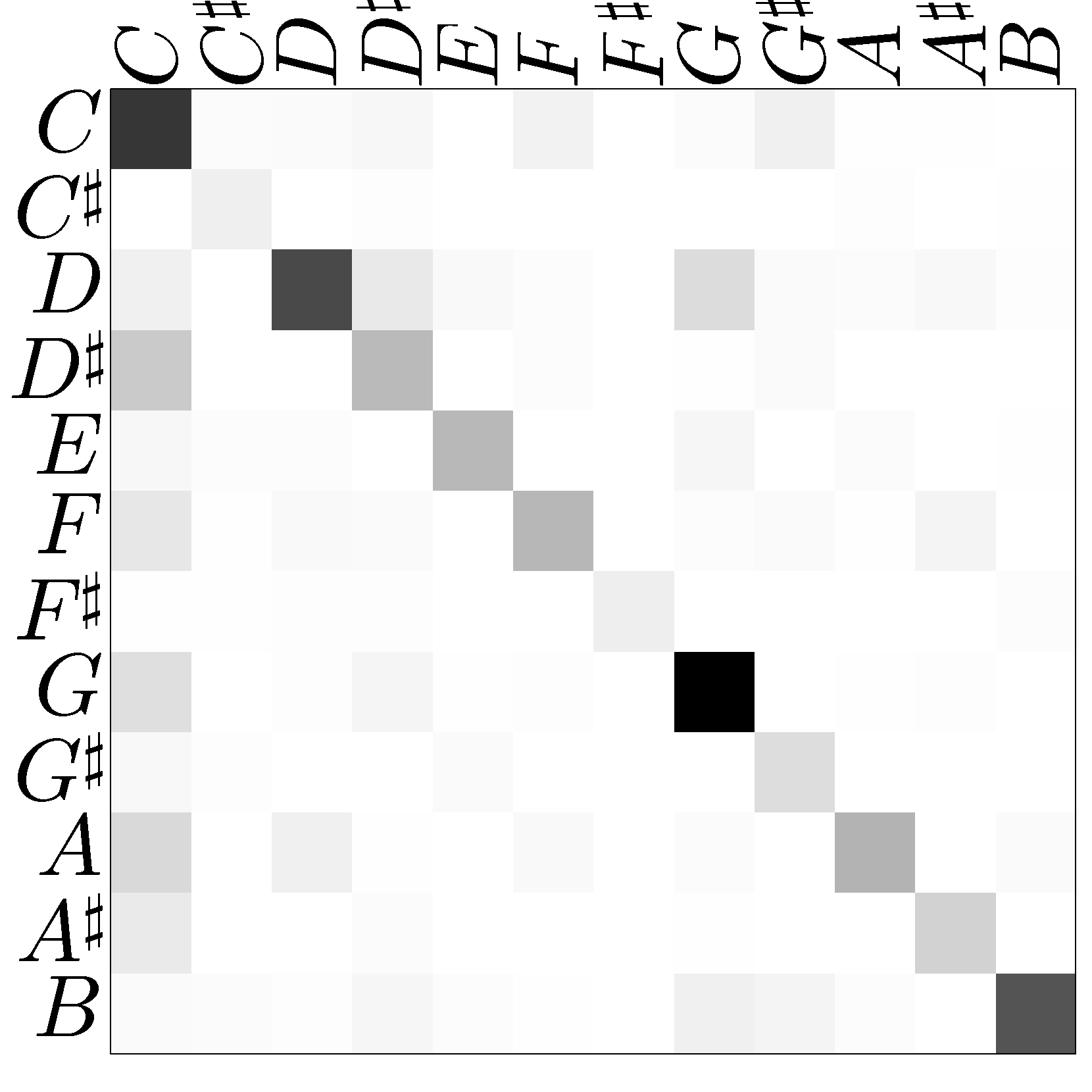}$\ \,$ &
\includegraphics[width=0.235\linewidth]{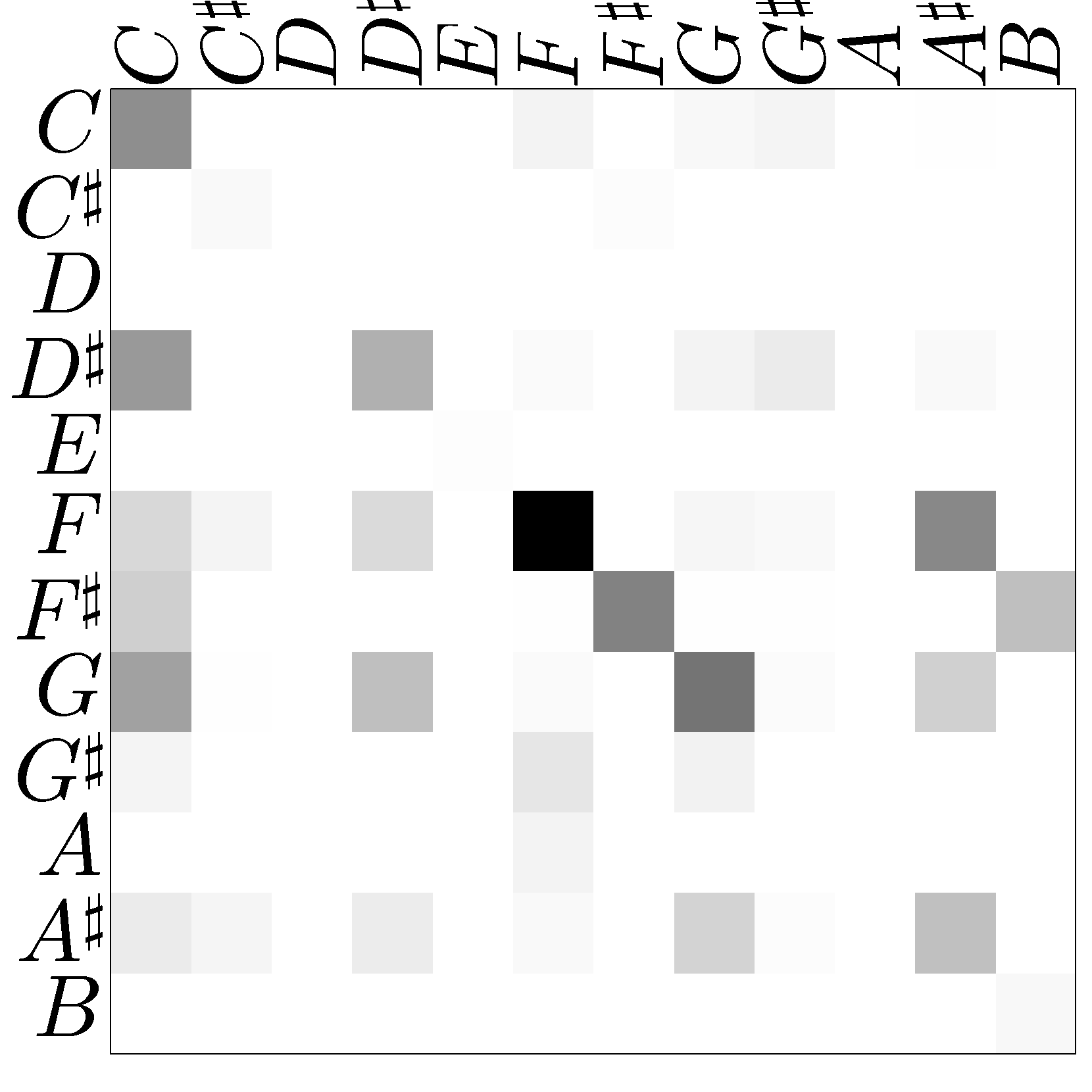}\\
(a) & (b) & (c) & (d) \\
\end{tabular}
\caption{\small Co-occurrence of tones in melody (y-axis) and chord (x-axis). (a)-(d) shows \textit{Major}(\textit{Minor}), \textit{Harmonic Minor}, \textit{Melodic Minor}, and \textit{Blues}, respectively.}%
\label{fig:chord}%
\end{figure}

We studied all existing chords in our 100 hours of pop music. Although in principle a chord can be any arbitrary combination of multiple notes, we observed that in the actual music data $99.19\%$ of the chords belong to one of 72 chord classes (6 types $\times$ 12 start notes). 
Fig.~\ref{fig:chord} shows the correlation between the melody's tone and the starting note of the chord playing at the same time. It can be seen that chord is strongly correlated with melody. These two findings inspire our design.
We thus represent chord ${\bf y}^t_{chd}$ as a one-hot encoding with 72 classes, and predict it using a two-layer LSTM with a 512-dimensional hidden state. 
We generate one chord at each time step. The input is ${\bf y}^{t-4}_{chd}$ concatenated with ${\bf y}^{t-3:t}_{key}$.

We look at our music dataset and  find all unique drum patterns with duration of a half bar. We then compute the histogram of all the patterns. This forms a long tail distribution, where $94.60\%$ comes from the top 100 common patterns.
We generate drum conditioned on the key layer using a two-layer LSTM with 512 dimensional hidden states. Drum ${\bf y}^t_{drm}$ is represented as one-hot encoding with of the 100 unique one-bar-long drum patterns. The input is ${\bf y}^{t-4}_{drm}$ concatenated with the notes from the previous three times steps ${\bf y}^{t-3:t}_{key}$.



\subsection{Learning}

We use cross-entropy as our loss function to train each layer. 
We follow the typical training strategy where we make predictions at each layer and time step but feed in ground-truth information to the next. This effectively decomposes training, and allows to   train all layers in parallel. We use the Adam optimizer, a learning rate of 2e-3 and a learning rate decay of  0.99 after each epoch for 10 epochs. 

\subsection{Music Synthesis: Putting all the Outputs Together}

To synthesize music we first randomly choose a scale and a profile ${\bf x}_{prf}$. For generating ${\bf x}_{prf}$, we randomly choose one cluster id with a random duration, and repeat until we get the desired total length of the music sequence. We then perform inference in our model conditioned on the chosen scale, and use ${\bf x}_{prf}$ as input to our key layer. At each time step, we sample a key according to $P({\bf y}^t_{key})$. We encode it as a one-hot vector and pass to the press, chord and drum layers. We sample the press, chords and drums at each time step in a similar fashion.

Before putting the outputs across layers together, we further adjust the generated sequences at the bar level. For melody, we first check at each bar if the first step is a continuation of a previous note or silence. If it is the latter, we find the first newly pressed note within the bar and move it to the beginning of the bar. We  do  similarly for the windows of two half-bars  as well as the four quarter-bars. This makes the melody more likely to be on the beat, and generally sounds better. We verify this in our experiments.

For chord, we generate one chord at each half bar, which is the majority of all single step chord generations. Furthermore, we incorporate the rule of chord progression in the \textit{Circle of Fifths} as between chords pairwise smooth terms, and compute the final chord using dynamic programming. For drum, we generate one pattern at each half bar. 

Our model  generates with scale starting note $C$, and then applies a constant shift to generate music with other starting notes. 
Besides scale, which instrument to use is also customizable. However, we simply set all instruments as grand piano in all experiments, as the effect and musical meaning of different instrument combinations is beyond the scope of this paper. 






%% file: fig1.tex
\begin{centering}
\begin{tikzpicture}[ scale=0.75 ]

\node[text width=0.5cm](xtprf) at(10*1.5,0) {${\bf x}^{t}_{prf}$};
\node[circle, draw]     (ctkey) at(10*1.5,1) {};
\draw[->] (xtprf) --    (ctkey);
\node[text width=0.5cm](ytkey) at(10*1.5,2) {${\bf y}^{t}_{key}$};
\draw[->] (ctkey) --    (ytkey);
\node[circle, draw]     (ctprs) at(10*1.5,3) {};
\draw[->] (ytkey) --    (ctprs);
\node[text width=0.5cm](ytprs) at(10*1.5,4) {${\bf y}^{t}_{prs}$};
\draw[->] (ctprs) --    (ytprs);
\node[circle, draw]     (ctchd) at(10*1.5,5) {};
\node[text width=0.5cm](ytchd) at(10*1.5,6) {${\bf y}^{t}_{chd}$};
\draw[->] (ctchd) --    (ytchd);
\node[circle, draw]     (ctdrm) at(10*1.5,7) {};
\node[text width=0.5cm](ytdrm) at(10*1.5,8) {${\bf y}^{t}_{drm}$};
\draw[->] (ctdrm) --    (ytdrm);

\node[text width=0.5cm](x1prf) at(9*1.5,0) {${\bf x}^{t-1}_{prf}$};
\node[circle, draw]     (c1key) at(9*1.5,1) {};
\draw[->] (x1prf) --    (c1key);
\node[text width=0.5cm](y1key) at(9*1.5,2) {${\bf y}^{t-1}_{key}$};
\draw[->] (c1key) --    (y1key);
\node[circle, draw]     (c1prs) at(9*1.5,3) {};
\draw[->] (y1key) --    (c1prs);
\node[text width=0.5cm](y1prs) at(9*1.5,4) {${\bf y}^{t-1}_{prs}$};
\draw[->] (c1prs) --    (y1prs);

\node[text width=0.5cm](x2prf) at(8*1.5,0) {${\bf x}^{t-2}_{prf}$};
\node[circle, draw]     (c2key) at(8*1.5,1) {};
\draw[->] (x2prf) --    (c2key);
\node[text width=0.5cm](y2key) at(8*1.5,2) {${\bf y}^{t-2}_{key}$};
\draw[->] (c2key) --    (y2key);
\node[circle, draw]     (c2prs) at(8*1.5,3) {};
\draw[->] (y2key) --    (c2prs);
\node[text width=0.5cm](y2prs) at(8*1.5,4) {${\bf y}^{t-2}_{prs}$};
\draw[->] (c2prs) --    (y2prs);

\node[text width=0.5cm](x3prf) at(7*1.5,0) {${\bf x}^{t-3}_{prf}$};
\node[circle, draw]     (c3key) at(7*1.5,1) {};
\draw[->] (x3prf) --    (c3key);
\node[text width=0.5cm](y3key) at(7*1.5,2) {${\bf y}^{t-3}_{key}$};
\draw[->] (c3key) --    (y3key);
\node[circle, draw]     (c3prs) at(7*1.5,3) {};
\draw[->] (y3key) --    (c3prs);
\node[text width=0.5cm](y3prs) at(7*1.5,4) {${\bf y}^{t-3}_{prs}$};
\draw[->] (c3prs) --    (y3prs);

\node[text width=0.5cm](x4prf) at(6*1.5,0) {${\bf x}^{t-4}_{prf}$};
\node[circle, draw]     (c4key) at(6*1.5,1) {};
\draw[->] (x4prf) --    (c4key);
\node[text width=0.5cm](y4key) at(6*1.5,2) {${\bf y}^{t-4}_{key}$};
\draw[->] (c4key) --    (y4key);
\node[circle, draw]     (c4prs) at(6*1.5,3) {};
\draw[->] (y4key) --    (c4prs);
\node[text width=0.5cm](y4prs) at(6*1.5,4) {${\bf y}^{t-4}_{prs}$};
\draw[->] (c4prs) --    (y4prs);
\node[circle, draw]     (c4chd) at(6*1.5,5) {};
\node[text width=0.5cm](y4chd) at(6*1.5,6) {${\bf y}^{t-4}_{chd}$};
\draw[->] (c4chd) --    (y4chd);
\node[circle, draw]     (c4drm) at(6*1.5,7) {};
\node[text width=0.5cm](y4drm) at(6*1.5,8) {${\bf y}^{t-4}_{drm}$};
\draw[->] (c4drm) --    (y4drm);

\node[text width=0.5cm] at(5*1.5,0) {...};
\node[text width=0.5cm] at(5*1.5,1) {...};
\node[text width=0.5cm] at(5*1.5,2) {...};
\node[text width=0.5cm] at(5*1.5,3) {...};
\node[text width=0.5cm] at(5*1.5,4) {...};

\node[text width=0.5cm](x8prf) at(4*1.5,0) {${\bf x}^{t-8}_{prf}$};
\node[circle, draw]     (c8key) at(4*1.5,1) {};
\draw[->] (x8prf) --    (c8key);
\node[text width=0.5cm](y8key) at(4*1.5,2) {${\bf y}^{t-8}_{key}$};
\draw[->] (c8key) --    (y8key);
\node[circle, draw]     (c8prs) at(4*1.5,3) {};
\draw[->] (y8key) --    (c8prs);
\node[text width=0.5cm](y8prs) at(4*1.5,4) {${\bf y}^{t-8}_{prs}$};
\draw[->] (c8prs) --    (y8prs);
\node[circle, draw]     (c8chd) at(4*1.5,5) {};
\node[text width=0.5cm](y8chd) at(4*1.5,6) {${\bf y}^{t-8}_{chd}$};
\draw[->] (c8chd) --    (y8chd);
\node[circle, draw]     (c8drm) at(4*1.5,7) {};
\node[text width=0.5cm](y8drm) at(4*1.5,8) {${\bf y}^{t-8}_{drm}$};
\draw[->] (c8drm) --    (y8drm);

\node[text width=0.5cm](x9prf) at(3*1.5,0) {${\bf x}^{t-9}_{prf}$};
\node[circle, draw]     (c9key) at(3*1.5,1) {};
\draw[->] (x9prf) --    (c9key);
\node[text width=0.5cm](y9key) at(3*1.5,2) {${\bf y}^{t-9}_{key}$};
\draw[->] (c9key) --    (y9key);
\node[circle, draw]     (c9prs) at(3*1.5,3) {};
\draw[->] (y9key) --    (c9prs);
\node[text width=0.5cm](y9prs) at(3*1.5,4) {${\bf y}^{t-9}_{prs}$};
\draw[->] (c9prs) --    (y9prs);

\node[text width=0.5cm] at(2*1.5,0) {...};
\node[text width=0.5cm] at(2*1.5,1) {...};
\node[text width=0.5cm] at(2*1.5,2) {...};
\node[text width=0.5cm] at(2*1.5,3) {...};
\node[text width=0.5cm] at(2*1.5,4) {...};
\node[text width=0.5cm] at(2*1.5,5) {...};
\node[text width=0.5cm] at(2*1.5,6) {...};
\node[text width=0.5cm] at(2*1.5,7) {...};
\node[text width=0.5cm] at(2*1.5,8) {...};

\node[text width=0.5cm](x16prf) at(1*1.5,0) {${\bf x}^{t-16}_{prf}$};
\node[circle, draw]     (c16key) at(1*1.5,1) {};
\draw[->] (x16prf) --    (c16key);
\node[text width=0.5cm](y16key) at(1*1.5,2) {${\bf y}^{t-16}_{key}$};
\draw[->] (c16key) --    (y16key);
\node[circle, draw]     (c16prs) at(1*1.5,3) {};
\draw[->] (y16key) --    (c16prs);
\node[text width=0.5cm](y16prs) at(1*1.5,4) {${\bf y}^{t-16}_{prs}$};
\draw[->] (c16prs) --    (y16prs);
\node[circle, draw]     (c16chd) at(1*1.5,5) {};
\node[text width=0.5cm](y16chd) at(1*1.5,6) {${\bf y}^{t-16}_{chd}$};
\draw[->] (c16chd) --    (y16chd);
\node[circle, draw]     (c16drm) at(1*1.5,7) {};
\node[text width=0.5cm](y16drm) at(1*1.5,8) {${\bf y}^{t-16}_{drm}$};
\draw[->] (c16drm) --    (y16drm);

\node[text width=0.5cm](x17prf) at(0*1.5,0) {${\bf x}^{t-17}_{prf}$};
\node[circle, draw]     (c17key) at(0*1.5,1) {};
\draw[->] (x17prf) --    (c17key);
\node[text width=0.5cm](y17key) at(0*1.5,2) {${\bf y}^{t-17}_{key}$};
\draw[->] (c17key) --    (y17key);
\node[circle, draw]     (c17prs) at(0*1.5,3) {};
\draw[->] (y17key) --    (c17prs);
\node[text width=0.5cm](y17prs) at(0*1.5,4) {${\bf y}^{t-17}_{prs}$};
\draw[->] (c17prs) --    (y17prs);

\draw[->] (c1key) -- (ctkey);
\draw[->] (y1key) -- (ctkey);
\draw[->] (c1prs) -- (ctprs);
\draw[->] (y1prs) -- (ctprs);
\draw[->] (c2key) -- (c1key);
\draw[->] (y2key) -- (c1key);
\draw[->] (c2prs) -- (c1prs);
\draw[->] (y2prs) -- (c1prs);
\draw[->] (c3key) -- (c2key);
\draw[->] (y3key) -- (c2key);
\draw[->] (c3prs) -- (c2prs);
\draw[->] (y3prs) -- (c2prs);
\draw[->] (c4key) -- (c3key);
\draw[->] (y4key) -- (c3key);
\draw[->] (c4prs) -- (c3prs);
\draw[->] (y4prs) -- (c3prs);
\draw[->] (c9key) -- (c8key);
\draw[->] (y9key) -- (c8key);
\draw[->] (c9prs) -- (c8prs);
\draw[->] (y9prs) -- (c8prs);
\draw[->] (c17key) -- (c16key);
\draw[->] (y17key) -- (c16key);
\draw[->] (c17prs) -- (c16prs);
\draw[->] (y17prs) -- (c16prs);
\draw[->] (c4chd) -- (ctchd);
\draw[->] (y4chd) -- (ctchd);
\draw[->] (c4drm) -- (ctdrm);
\draw[->] (y4drm) -- (ctdrm);
\draw[->] (c8chd) -- (c4chd);
\draw[->] (y8chd) -- (c4chd);
\draw[->] (c8drm) -- (c4drm);
\draw[->] (y8drm) -- (c4drm);

\draw[blue,->] (ytkey) .. controls(9.5*1.5,4) .. (ctchd);
\draw[blue,->] (ytkey) .. controls(9.5*1.5,5) .. (ctdrm);
\draw[blue,->] (y1key) .. controls(8.5*1.5,4) .. (ctchd);
\draw[blue,->] (y1key) .. controls(8.5*1.5,5) .. (ctdrm);
\draw[blue,->] (y2key) .. controls(7.5*1.5,4) .. (ctchd);
\draw[blue,->] (y2key) .. controls(7.5*1.5,5) .. (ctdrm);
\draw[blue,->] (y3key) .. controls(6.5*1.5,4) .. (ctchd);
\draw[blue,->] (y3key) .. controls(6.5*1.5,5) .. (ctdrm);
\draw[blue,->] (y4key) .. controls(5.5*1.5,4) .. (c4chd);
\draw[blue,->] (y4key) .. controls(5.5*1.5,5) .. (c4drm);
\draw[blue,->] (y8key) .. controls(3.5*1.5,4) .. (c8chd);
\draw[blue,->] (y8key) .. controls(3.5*1.5,5) .. (c8drm);
\draw[blue,->] (y9key) .. controls(2.5*1.5,4) .. (c8chd);
\draw[blue,->] (y9key) .. controls(2.5*1.5,5) .. (c8drm);
\draw[blue,->] (y16key) .. controls(0.5*1.5,4) .. (c16chd);
\draw[blue,->] (y16key) .. controls(0.5*1.5,5) .. (c16drm);
\draw[blue,->] (y17key) .. controls(-0.5*1.5,4) .. (c16chd);
\draw[blue,->] (y17key) .. controls(-0.5*1.5,5) .. (c16drm);

\draw[red,->] (y8key) .. controls(6*1.5,0.2) .. (ctkey);
\draw[red,->] (y9key) .. controls(6*1.5,0.2) .. (ctkey);
\draw[red,->] (y16key) .. controls(3*1.5,0) .. (ctkey);
\draw[red,->] (y17key) .. controls(3*1.5,0) .. (ctkey);

\draw[dashed] (-1,0.6) -- (18,0.6) -- (18,2.35) -- (-1,2.35) -- cycle;
\draw[dashed] (-1,2.6) -- (18,2.6) -- (18,4.35) -- (-1,4.35) -- cycle;
\draw[dashed] (-1,4.6) -- (18,4.6) -- (18,6.35) -- (-1,6.35) -- cycle;
\draw[dashed] (-1,6.6) -- (18,6.6) -- (18,8.35) -- (-1,8.35) -- cycle;
\node[text width=3cm, align=right](namekey) at(15.9,0.85) {Key Layer$|s$};
\node[text width=2cm, align=right](nameprs) at(16.6,2.85) {Press Layer};
\node[text width=2cm, align=right](namechd) at(16.6,4.85) {Chord Layer};
\node[text width=2cm, align=right](namedrm) at(16.6,6.85) {Drum Layer};
\end{tikzpicture}
\end{centering}

%% file: iclr-experiment.tex
\section{Experiments}

To train our model, we took 100 hours of pop music from~\cite{midiman} which consists of user-composed pop songs and video game music. In our generation, we always use 120 beats per minute with 4 time steps per beat. However, songs in the dataset can have arbitrary speed. To neutralize the effect of this, we detect the most frequent interval between two adjacent notes for each song, and iteratively divide or multiply this interval by 2 until it falls in the range between $0.25s$ and $0.5s$. We use this as a measure of the song's beat duration. We then adjust the song's temporal axis so that all songs have the same beat duration of $0.5s$.

\begin{figure}[t]%
\centering
\includegraphics[width=0.8\linewidth]{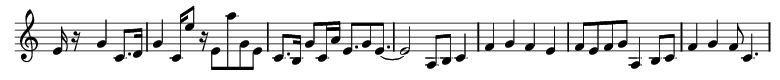}\\
\includegraphics[width=0.8\linewidth]{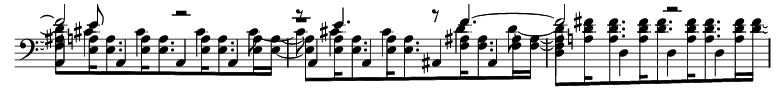}\\
\includegraphics[width=0.8\linewidth]{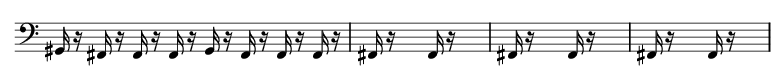}\\
\caption{{\small Example of our music generation. From top to bottom: melody, chord and drum respectively.}}%
\label{fig:sheet}%
\end{figure}

A MIDI file can be separated into different channels/tracks, where the 9th channel is specifically preserved for drums. We categorize the rest of non-drum tracks into melody, chord, and else, by simply setting thresholds on average number of unique notes within a bar and average number of note changing within a bar, as chords are by definition repetitive. Fig.~\ref{fig:sheet} shows an example of our music generation.

To evaluate the quality of our music generation, we conduct a human survey with 27 participants. In the survey, participants are presented with several pairs of 30-second music clips, and are asked to vote which clip in the pair sounds better. We gave no other information about what they are listening to. They are also allow to submit a neutral vote in case they cannot decide between the two choices. In our study, we consider three cases: our full method versus Magenta~\cite{magenta}, our method with melody only versus Google Magenta (\cite{magenta}), and our method versus our method without the temporal alignment described in Sec.4.5. We randomly generated 10 songs per method and  randomly shuffled each pair. 

As shown in  Table~\ref{tab:human}, most participants prefer songs produced by our method compared to Magenta. Participants also made comments such as \textit{music sounds better with percussion than piano alone}, and \textit{multiple instruments with continuous play is much better}. This confirms that our multi-layer generation improves music quality. Few participants also point out that \textit{drums sound too different and do not participate to the melody perfectly}, which indicates that further improvements can be still made. In the second comparison, we study if the quality improvement of our method is only caused by adding chords and drums, or is also related to our two-layer melody generation with alignment. It can be seen that without chords and drums, the score drops as expected, but is still much higher than the Magenta baseline. This is because our method \textit{produces less recursion and silence}, and \textit{faster and more accurate tempo} as mentioned by the participants. In the last comparison, most participants prefer our full method than the no-alignment version, since \textit{beats are more subtle and better timed}. This proves the usefulness of temporal alignment.

Finally we study our model's capabilities to generate new music.  
Towards this goal, we generated 100  sequences of 50 seconds of length using different random initializations. 
Then for each sequence, we search for the longest sub-sequence of keys that matches part of the training data, and record its length. We find out that with 1 hour of training data, the mean matching sub-sequence length is $3.46s$. With 100 hours of training data, the mean length increases to $4.65s$, since there are more possible matches. 
The sequences are very small and thus, our model is able to generate new music. 

\begin{table}
\begin{center}
\vspace{-1mm}
\addtolength{\tabcolsep}{3pt}
\begin{tabular}{|c||c|c||c|c||c|c|}
\hline
Method		& Ours	& Magenta	& Ours-MO	& Magenta	& Ours	& Ours-NA\\
\hline
\hline
\% of votes	& $81.6\%$	& $14.4\%$	& $69.6\%$	& $13.6\%$	& $75.2\%$	& $12.0\%$\\
\hline
\end{tabular}
\end{center}
\caption{\small Human evaluation of music generated by different methods: ours and~\cite{magenta}'s Magenta. Ours-MO and Ours-NA are short for Ours Melody Only and Ours No Alignment. We allowed neutral votes, thus the sum of the pair is less than 100\%. }
\label{tab:human}
\end{table}

%% file: iclr-application.tex
\section{Applications}

In this section we demonstrate two novel applications of our pop music generation framework. 
We refer the reader to {\color{magenta} http://www.cs.toronto.edu/songfrompi/ } for the music videos. 

\subsection{Neural Dancing and Karaoke}

\begin{figure}[t!]%
\centering
\addtolength{\tabcolsep}{-8pt}
\begin{tabular}{cccc}
\includegraphics[width=0.235\linewidth]{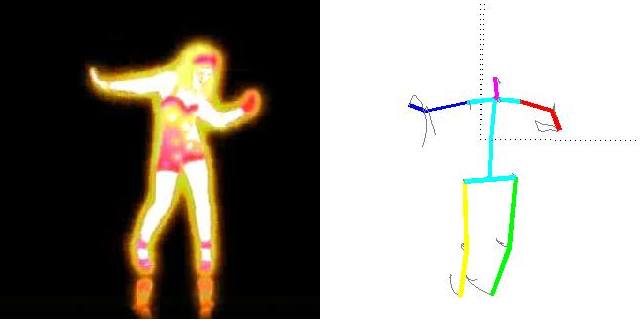}$\ \,$ & 
\includegraphics[width=0.235\linewidth]{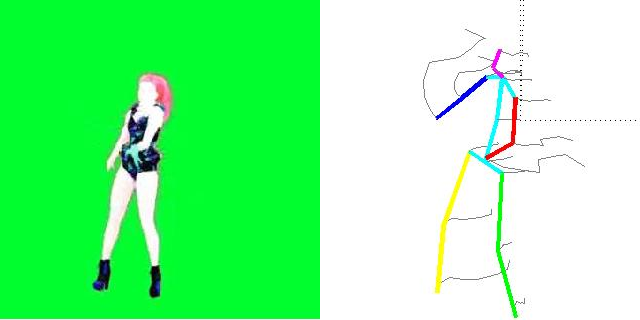}$\ \,$ &
\includegraphics[width=0.235\linewidth]{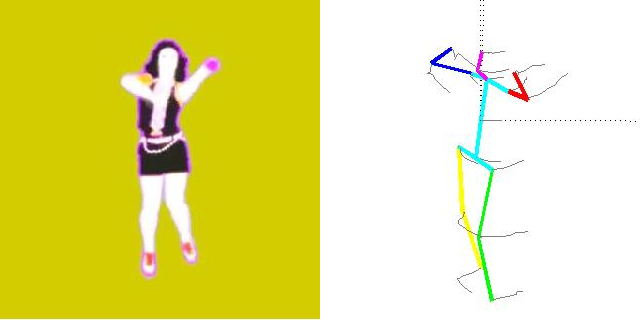}$\ \,$ &
\includegraphics[width=0.235\linewidth]{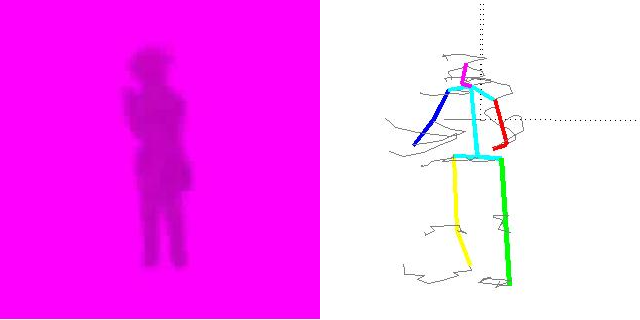}\\
(a) & (b) & (c) & (d) \\
\end{tabular}
\caption{\small Examples from \emph{Just Dance} and 3D human pose tracking result. (a) and (b) are success cases, pose tracking fails in (c), and (d) shows the defect in video which makes tracking difficult.}%
\label{fig:dance}%
\end{figure}

In our first application, we attempt to generate both music and a stickman dancing to it, as well as a sequence of karaoke-like text that people can sing along with. 
To learn the relationship between music and dance, we download 1 hour of video from the game \textit{Just Dance}, as well as the MIDI files for songs included in the video from different sources. We use the method in~\cite{newell2016stacked} to track single-frame 2D human pose in the videos. We process the single-frame tracking result to ensure left-right body  consistency through time, and then use the method of~\cite{zhou2015sparseness} to convert the 2D pose sequence into 3D. Example results are shown in Fig.~\ref{fig:dance}. We observe that our pose processing pipeline is able to extract reasonable human poses most of the time. However, the quality is not perfect due to tracking failure or video effects. We define pose similarity as average euclidean distance of all joints, and cluster poses into 456 clusters. We used~\cite{frey2007clustering} as the number of clusters is large.

We learn to generate a stickman dancing by adding another dancing layer on top of the key layer, just like for drum and chord. We generate one pose at each beat, which is equivalent to 4 time steps or 0.5 seconds in a 120 beat-per-minute music. In particular, we predict one of the 456 pose clusters using a linear projection layer followed by softmax.  We use cross-entropy at each time step as our loss function. At inference time, we further apply moving average to temporally smooth the generated 3D pose sequence.

To learn the relationship between music and lyrics, we collect 51 hours of lyrics data from the internet. This data contains 50 hours of text without music, and the rest 1 hour are songs we collected from \textit{Just Dance}. For the music part, we temporally align each sentence in the lyrics with the midi music by using the widely-existing \emph{lrc} format, which records the time tag at the beginning of every sentence. We select words that appear at least 4 times, which yields a vocabulary size of 3390 including unknown and end-of-sentence. Just as for dance, we generate one word per beat using another lyrics layer on top of the key layer.

\subsection{Neural Story Singing}

In this application our aim is to sing a song about a photo. We first generate a story about the photo with the neural storyteller~\cite{kiros2015skip} and try to accompany the generated text with music.
We utilize the same 1 hour dataset of temporally aligned lyrics and music. We further include the phoneme list of our 3390 vocabulary as we also want to sing the story. Starting from the text produced by neural storyteller, we arrange it into a temporal sequence with 1 beat per word and a short pause for end-of-sentence, where the pause length is decided such that the next sentence starts from a new bar. As our dataset is relatively small, we generate the profile conditioned on the text, which has less dimensions compared to the key. This is done by a 2-layer LSTM that takes as input the generated profile at the last time step concatenated with a one-hot vector of the current word, and outputs the current profile. We then generate the song with our model given the generated profile. The generated melody key is then used to decide on the pitch frequency of a virtual singer, assuming the key-to-pitch correspondence of a grand piano. We further constrain that the singer's final pitch is always in the range of $E3$ to $G4$, which we empirically found to be the natural pitch range. We then replace all words outside the vocabulary with the sound \textit{Ooh}, and play the rendered singing with the generated music.


\section{Conclusion and Future Work}

We have presented a hierarchical approach to pop song generation which exploits music theory in the model design. In contrast to past work, our approach is able to generate multi-track music. Our human studies shows the strength of our framework compared to an existing strong baseline. We additionally proposed two new applications: neural dancing \& karaoke, and neural story singing. 
We next discuss the limitations and avenues for future work. As most existing approaches our method's objective is to learn to produce music at the note level. This can be unsuitable for music, as music is flexible and intentionally made to be unpredictable when it is composed. This calls for a deeper study of music theory, as in this paper we are only  
scratching the surface.